\documentclass[amssymb,pra,twocolumn,notitlepage,superscriptaddress]{revtex4-1}

\renewcommand{\vec}[1]{\mathbf{#1}}

\newcommand{\braket}[2]{\langle #1 | #2 \rangle}

\newcommand{\ket}[1]{| #1 \rangle}
\newcommand{\bra}[1]{\langle #1 |}

\usepackage{color}

\usepackage{amsmath}                          
\usepackage{amsfonts}                         
\usepackage{amssymb}                          
\usepackage{amsthm}                           
\usepackage{mathdots}                         

\usepackage{graphicx}                         
\usepackage{subfigure}                        
\usepackage{overpic}

\begin{document}


\title{Realizing quantum linear regression with auxiliary qumodes}
\author{Dan-Bo Zhang}
\affiliation{Guangdong Provincial Key Laboratory of Quantum Engineering
    and Quantum Materials, and School of Physics\\ and Telecommunication Engineering,
    South China Normal University, Guangzhou 510006, China}

\author{Zheng-Yuan Xue}
\affiliation{Guangdong Provincial Key Laboratory of Quantum Engineering
    and Quantum Materials, and School of Physics\\ and Telecommunication Engineering,
    South China Normal University, Guangzhou 510006, China}

\author{Shi-Liang Zhu} \email{slzhu@nju.edu.cn}
\affiliation{National Laboratory of Solid State Microstructures,
School of Physics, Nanjing University, Nanjing 210093, China}
\affiliation{Guangdong Provincial Key Laboratory of Quantum
Engineering
    and Quantum Materials, and School of Physics\\ and Telecommunication Engineering,
    South China Normal University, Guangzhou 510006, China}

\author{Z. D. Wang}
\email{zwang@hku.hk} \affiliation{Department of Physics and
Center of Theoretical and Computational Physics, The University of
Hong Kong, Pokfulam Road, Hong Kong, China}

\begin{abstract}
In order to exploit quantum advantages, quantum algorithms are
indispensable for operating machine learning with quantum
computers. We here propose an intriguing hybrid approach of
quantum information processing for quantum linear regression,
which utilizes both discrete  and continuous quantum variables, in
contrast to existing wisdoms based solely upon discrete qubits. In
our framework, data information is encoded in a qubit system,
while information processing is tackled using auxiliary continuous
qumodes via qubit-qumode interactions. Moreover, it is also
elaborated that finite squeezing is quite helpful for efficiently
running the quantum algorithms in realistic setup. Comparing with
an all-qubit approach, the present hybrid approach is more
efficient and feasible for implementing quantum algorithms, still
retaining exponential quantum speed-up.

\end{abstract}


\maketitle

\section{Introduction}
For quantum systems, there exist both discrete and continuous
variables, and they may  interact with each other. A typical
example is an light-atom interacting system, which demonstrates
how discrete energy levels and continuous light fields evolve
quantum mechanically. Notably, a hybrid scenario of information
processing using both discrete and continuous variables has been
proposed
before~\cite{furusawa2011quantum,andersen_15,lloyd00,van_08,proctor_17},
such as hybrid quantum computing~\cite{lloyd00,van_08,proctor_17}.
The relevant advantages have been indicated in tasks like quantum
float computing~\cite{lloyd_99} and quantum phase
estimation(QPE)~\cite{lloyd00,liu_16}, in which  infinite
dimensions of continuous variables are
exploited~\cite{lau_16,lau_17,aaronson_15,liu_16}, making the
proposals promising.


As is known, machine learning plays an important role for
extracting worthwhile information and  making trustable
predictions in an era of big
data~\cite{christopher_m_bishop_pattern_2006}. In addition, a
magnificent combination of machine learning and quantum mechanics
has opened a new window for information
processing~\cite{biamonte_17,wiebe_12,schuld_15,lloyd_14,rebentrost_14,dunjko_16,lloyd_16,wang_17}.
A class of quantum machine
learning~\cite{harrow_09,clader_13,pan_14,wang_17} is based on the
Harrow-Hassidim -Lloyd (HHL) algorithm that aims to obtain the
inverse of a matrix with exponential speed-up under reasonable
conditions~\cite{harrow_09}.  Normally, quantum linear regression
has been regarded as a representative task in quantum machine
learning and investigated by various all-qubit
approaches~\cite{wiebe_12,schuld_16,wang_17}. The algorithms for
quantum linear regression usually consist of several main parts,
namely quantum phase estimation, regularization, singular value
transformation, and prediction. All of these approaches require
ancillary qubits to register singular values in quantum phase
estimations that are necessary for quantum linear regression.
Depending on the desired precision, the number of ancillary qubits
may be large, which demands a more rare resource of qubits and is
a grand challenge for current quantum technology. On the other
hand, the quantum phase estimation can be implemented much more
efficiently using the hybrid approach~\cite{lloyd00}, which
requires merely one qumode based on appropriate squeezing
states~\cite{liu_16}. This motivates us to work out a quantum
algorithm of linear regression using the hybrid approach with
desired quantum advantages.

In this paper, we first introduce how to convert a linear
regression into a quantum task.  We then recall the quantum
algorithms of all-qubit systems, and analyze their properties. We
emphasize the demanding of ancillary qubits, and the inefficiency
of incorporating regularization and realizing singular value
transformation. By introducing qumodes, we propose a hybrid
approach for quantum linear regression, where single values are
encoded into the entangled two qumodes and single value
transformation can be implemented simply by homodyne measurements
with post-selection. This makes the algorithm more feasible for future physical implementation as it involves basic continuous variable quantum operations, in contrast to the all-qubit approach that requires complicated quantum circuits for quantum arithmetic and control-rotation.  A brief proposal for a physical realization
of the algorithm with trapped ions is suggested. Our results show
that the hybrid approach still retains the same order of runtime
$O(\log{MN})$ as the all-qubit approach. The regularization can be
incorporated by a controlled-phase gate on two qumodes, and it may
greatly reduce the required squeezing factor for the case of bad
condition number~\cite{harrow_09,aaronson_15}. We also investigate
a basic role of the finite squeezing factor, and find that it is
not only helpful for running the algorithm efficiently, but also
may be taken as an extra regularization for regression.

The paper is organized as follows. We introduce quantum linear regression in Sec.~\ref{Sec: quantum_version}, and give an analysis of the existing all-qubit approach in Sec.~\ref{Sec: existing_approach}. Then, in Sec.~\ref{Sec: algorithm} we present the quantum algorithm using qumodes and its physical implementation in trapped ions. Finally, discussions and conclusions are given in Sec.~\ref{Sec: Discussion}.

\section{Quantum version of linear regression} \label{Sec: quantum_version}
In this section, we will demonstrate how to formulate the linear
regression as a  quantum problem that may be solved using quantum
algorithm. Let us first introduce the linear regression. Given a
training dataset of $M$ points $\{\mathbf{a}^{(m)},y^{(m)}\}$,
where $\mathbf{a}^{(m)} \in R^N$ is a vector of $N$ features and
$y^{(m)}\in R$ is the target value, the goal is to learn a linear
model with parameters $\mathbf{w} \in R^N$  that can give the
prediction $\tilde{y}$ for new data $\tilde{\mathbf{a}}$ as
$\tilde{y}=\tilde{\mathbf{a}}^T\mathbf{w}$. The parameters
$\mathbf{w}$ can be estimated by minimizing the loss function of
least-square errors 
\begin{equation}
L(\mathbf{w})=\sum_{m=1}^{M}(\mathbf{w}^T\mathbf{a}^{(m)}-y^{(m)})^2+\chi||\mathbf{w}||^2
\end{equation}
over $\mathbf{w}$. Here $\chi||\mathbf{w}||^2$ is a regularization term with
parameter $\chi$, which is usually  considered in machine learning
for better performance. It will be shown later that the
regularization is helpful for efficiently implementing the quantum
algorithm.
Introducing the matrix notation $A_{mn}=\mathbf{a}^{(m)}_n$,  the linear regression solution turns out to be $\mathbf{w}=A^+\mathbf{y}$, where
the Moore-Penrose pseudoinverse (with $\chi$ term) reads $A^+=(A^TA+\chi I)^{-1}A^T$. Here $I$ is the identity matrix.  
It is inspiring to study the linear regression by using the singular value decomposition of $A$ by \cite{schuld_16}, $A = \sum_i \lambda_i \mathbf{u}_i\mathbf{v}_i^T$.
Here $\lambda_i$ are singular values of $A$ with corresponding left (right) eigenvector $\mathbf{v}_i$ ($\mathbf{u}_i$). Now it can be verified that the Moore-Penrose pseudoinverse reads as
$A^+ = \sum_i \frac{\lambda_i}{\lambda_i^2+\chi} \mathbf{v}_i\mathbf{u}_i^T$. Then, the prediction can be written as
$\tilde{y} = \sum_i \frac{\lambda_i}{\lambda_i^2+\chi} \mathbf{u}_i^T\mathbf{y}\tilde{\mathbf{a}}^T\mathbf{v}_i$.

To formulate a quantum version of linear regression, we need to
encode each component as a quantum state or a quantum operator.
For a vector $\mathbf{x}=(x_1,x_2,...,x_N)$, the amplitude
encoding scheme is taken to encode $\mathbf{x}$ into a quantum
state  as $\ket{\psi_{\mathbf{x}}}=\sum_n x_n\ket{n}$. Here
$\ket{n}=\ket{n_1n_2n_3...}$,where $n=n_1n_2n_3...$ is a binary
representation of integer $n$, and it requires a number of
$[log_2N]$ qubits.  Accordingly, $\mathbf{y}$ and
$\tilde{\mathbf{a}}$ are encoded as
$\ket{\psi_{\mathbf{y}}}=\sum_m y^{(m)}\ket{m}$ and
$\ket{\psi_{\tilde{\mathbf{a}}}}=\sum_n\tilde{\mathbf{a}}_n\ket{n}$,
respectively. Without loss of generality, we assume all quantum
states in this paper are normalized, and the final result should
be rescaled accordingly.  The remaining question now is how to
treat $A^+$. It is natural to take it as a operator, and the
prediction finally writes as
$\tilde{y}=\bra{\psi_{\tilde{\mathbf{a}}}}A^+\ket{\psi_\mathbf{y}}$.
One should pay special attention to the following aspects for this
approach\cite{wiebe_12}. Firstly, $A^T$ is not necessary a square
matrix, and the Hilbert space should be extended to define a
square matrix that contains $A^T$.  Secondly, two  HHL-like
algorithms for applying $A^T$ and $(A^TA+\chi I)^{-1}$
sequentially are required. Thirdly, both $A^T$ and $(A^TA+\chi
I)^{-1}$ are not unitary in general, there must be some
non-unitary procedure such as measurements or projections in the
algorithm. Thus, measurement is required in the middle of the
algorithm.


An alternative approach to circumvent the above problems is to
treat $A^+$ as a quantum state~\cite{schuld_16}.  This is possible
by rewriting the prediction as a tensor formula
\begin{equation}
\tilde{y} = \sum_i \frac{\lambda_i}{\lambda_i^2+\chi}(\mathbf{u}_i\otimes \mathbf{v}_i)^T \mathbf{y}\otimes\tilde{\mathbf{a}}, \nonumber
\end{equation}
which is an inner product between two vectors. Accordingly, we may
write the quantum state corresponds to $A^+$  as $\ket{\psi_{A^+}}
=\sum_i
\frac{c\lambda_i}{\lambda_i^2+\chi}\ket{\psi_{\mathbf{u}_i}}\ket{\psi_{\mathbf{v}_i}}$,
where $c$ is introduced for normalization. Then, the prediction
$\tilde{y}$ is obtained as  the inner product between
$\ket{\psi_{A^+}}$ and
$\ket{\psi_{\mathbf{y}}}\otimes\ket{\psi_{\tilde{\mathbf{a}}}}$,
up to a constant factor $\frac{1}{c}$.

On the other hand, the input data $A$ can be loaded as a quantum
state
$\ket{\psi_{A}}=\sum_{m,n}\mathbf{a}^{(m)}_n\ket{m}\ket{n}=\sum_i
\lambda_i\ket{\psi_{\mathbf{u}_i}}\ket{\psi_{\mathbf{v}_i}}$,
where the second equality comes from Schmidt decomposition. Thus,
the key task is to find a  quantum algorithm that transforms
$\ket{\psi_{A}}$ to $\ket{\psi_{A^+}}$.  After loading data $A$ as
initial state $\ket{\psi_{A}}$, different quantum techniques may
be used to steer the system from initial state $\ket{\psi_{A}}$ to
the target state $\ket{\psi_{A^+}}$.


Since both $\ket{\psi_{A}}$ and $\ket{\psi_{A^+}}$ are bipartite
quantum states, an investigation  of their entanglement structure
may inspire the design of the quantum algorithm. They share the
same entangling basis of
$\{\ket{\psi_{\mathbf{u}_i}}\ket{\psi_{\mathbf{v}_i}}\}$ with
different coefficients, where $\ket{\psi_{\mathbf{v}_i}}\in R^N$
are eigenstates in feature space and $\ket{\psi_{\mathbf{u}_i}}\in
R^M$ are eigenstates related to sample space. Then a quantum
algorithm can be devised that keeps those basis unchanged, while
the coefficients are transformed in the form of
$g(\lambda)=\frac{\lambda}{\lambda^2+\chi}$. Such a transformation
has been realized using ancillary qubits
\cite{harrow_09,wiebe_12,schuld_16}.

It should be pointed out that focusing on $\ket{\psi_{A^+}}$
instead of the states of parameters $\ket{\psi_\mathbf{w}}$, which is defined as $\ket{\psi_\mathbf{w}}=\sum_n w_n\ket{n}$ for parameters $\vec{w}=(w_1,...,w_n)$, has
its own advantages, especially when there are multiple target
values for linear regression. For example, based on the features
of one person, one can predict two target values such as both
income and cost.  In such a case, predictions are obtained by
inner product of $\ket{\psi_{A^+}}$ and
$\ket{\psi_{\tilde{\mathbf{a}}}}\ket{\psi_{\mathbf{y}_l}}$($l=1,2$
for income and cost), separately. In other words, once
$\ket{\psi_{A^+}}$ is obtained it can be applied for linear
regression with different target vectors. On the other hand,
$\ket{\psi_{\mathbf{w}_t}}$ is specified to fixed target value.
Moreover, one may construct the later from the former for each
target vector $\ket{\psi_{\mathbf{y}_t}}$ as
\begin{equation}
\ket{\psi_{\mathbf{w}_t}}=\sum_i\frac{c\lambda_i}{\lambda_i^2+\chi} \braket{\psi_{\mathbf{y}_t}}{\psi_{\mathbf{u}_i}}\ket{\psi_{\mathbf{v}_i}}.
\end{equation}
In this sense, $\ket{\psi_{A^+}}$ is more fundamental than $\ket{\psi_{\mathbf{w}_t}}$. In our implementation, we adopt this more efficient way.

\section{Existing all-qubit approach} \label{Sec: existing_approach}
Before proposing the hybrid approach using both qubits and qumodes, we first recall how to do quantum linear regression with all-qubit systems, where regularization has not been considered~\cite{wiebe_12,schuld_16,wang_17}. An analysis is given to show that the all-qubit approach demands for lots of ancillary qubits in several main procedures of quantum linear regression.  This motives us to propose a hybrid approach that is more efficient.

Basically, there are two subroutines in the algorithm of all-qubit
approach.  Firstly, quantum phase estimation is applied that
registers $\lambda_i$ in ancillary qubits; secondly, coefficients
$\frac{1}{\lambda_i}$ are obtained by a conditional rotation on an
extra ancillary qubit and then a projection to $\ket{1}$ state.
The success rate of projecting to $\ket{1}$ is proportional to the
conditional number
$\kappa=\frac{\lambda^2_{\text{max}}}{\lambda^2_{\text{min}}}$,
where $\lambda_{\text{max/min}}$ stands for largest/smallest
singular values of $A$. Under bad conditional number, e.g.,
$\kappa_c\sim O(N)$ (which naturally arise for low-rank $A^TA$),
the runtime will scale with $O(N)$ and thus destroy the
exponentially speed-up. This issue, as will be demonstrated later,
can be remedied by introducing regularization.

We analyze the cost of resource, especially on the number of
qubits.  For the quantum phase estimation, firstly, series of
$\{e^{iA^TA\tau}\}$ with different time $\tau$ should be
constructed. Secondly, since $\lambda_i$ are typically continuous
numbers, to encode them with precision $\varepsilon$ a number of
$O(\log{\varepsilon^{-1}})$ qubits is required. If regularization
is included, a quantum circuit for quantum addition is required,
which requires extra $O(\log{\varepsilon^{-1}})$ qubits. For
singular value transformation, to realize the required conditional
rotation an oracle should be constructed which would involve many
qubits depending on the precision. Importantly, for all processes,
although lots of ancillary qubits should be used, they would be
discarded finally. Since their role is just to register singular
values and further give singular value transformation, it is
desirable for alternative approach that requires less resource of
qubits.

It is known that the quantum phase estimation as well as the
quantum addition are more naturally and  efficiently implemented
in a quantum system with continuous variables \cite{lloyd_99}
(also see Appendix  ~\ref{appendix: qumode}). On the other hand,
vectors of classical data are better encoded in quantum states of
qubit systems, as each component of a vector now becomes the
corresponding amplitude directly, while encoding in qumodes this
would be much complicated~\cite{lau_16,lau_17}. Given the above
two considerations, it naturally calls for a hybrid quantum
computing approach that exploits both the advantages of qubits and
qumodes, as would be explored in this paper.

\begin{figure}
    \includegraphics[width=0.8\columnwidth]{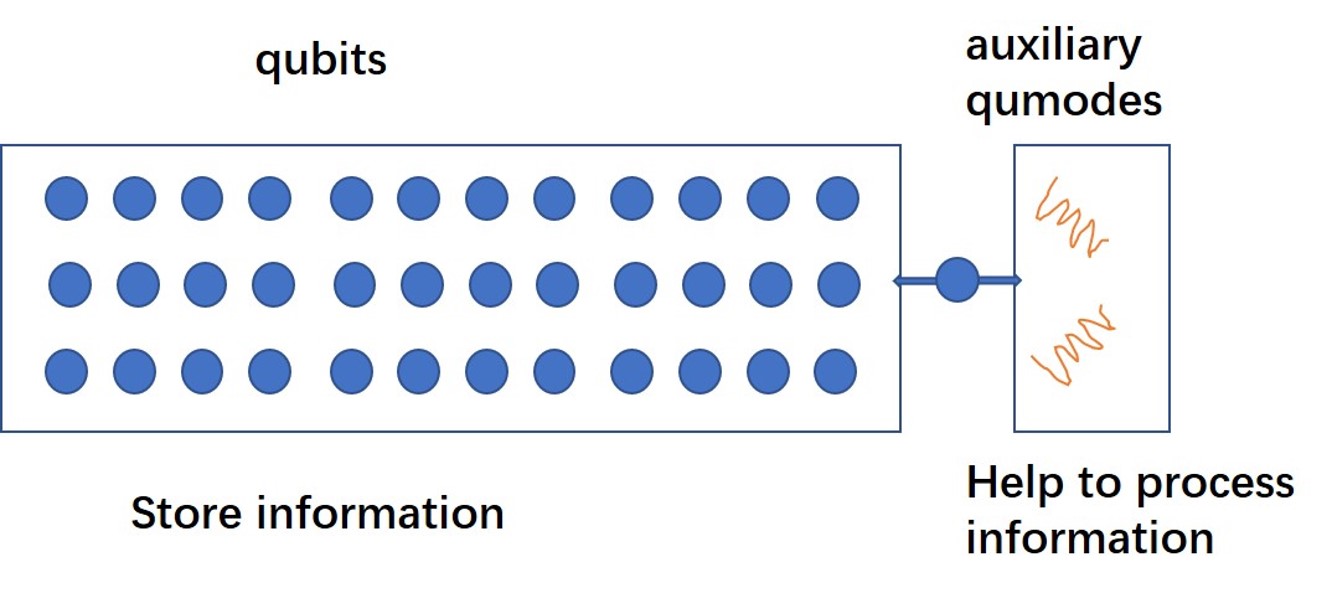}
    \caption{Illustration of the hybrid approach that exploits the best of both   qubits and qumodes. In our implementation of quantum linear regression qubits and qumodes are minimal coupled by connecting only to an ancillary qubit.}
\end{figure}


\section{Quantum algorithm using two auxiliary qumodes}\label{Sec: algorithm}
We present the quantum algorithm for linear regression using two
auxiliary qumodes.   Further discussions will be presented in
Sec.~\ref{Sec: Discussion}. We introduce qumodes $\ket{x}_q$ and
$\ket{x}_p$, which are eigenstates of conjugate quadrature field
operators $\hat{q}$ and $\hat{p}$ (such as momentum and position
operators), respectively. As $[\hat{q},\hat{p}]=i$, we have
$\ket{q}_q=\frac{1}{\sqrt{2\pi}}\int dpe^{-iqp}\ket{p}_p$. It
should be emphasized that the squeezing factor can not be infinite
physically. Nevertheless, we first present the algorithm for the
case of infinite squeezing, as the main procedures of the
algorithm is more clearly revealed. We then stress the importance
of finite squeezing and modify the algorithm accordingly.

The algorithm under infinite squeezing can be summarized as following:
\begin{enumerate}
    \item \emph{State preparation}. Load training data $\{\vec{a}^{(m)}\}$ into  a quantum state
$\ket{\psi_{A}}=\sum_m\ket{m}\ket{\psi_{\vec{a}^{(m)}}}$
with quantum random access
memory~\cite{giovannetti_08,giovannetti_08,hong_12,patton_13}. Two
qumodes are initialized in $\ket{0}_{q_1}\ket{0}_{q_2}$.
    \item \emph{Quantum phase estimation}. Perform $U=e^{i\eta A^TA\hat{p_1}\hat{p_2}}$ on $\ket{\psi_A}\ket{0}_{q_1}\ket{0}_{q_2}$ which leads to
    \[ \sum_i \lambda_i \ket{\psi_{\mathbf{u}_i}}\ket{\psi_{\mathbf{v}_i}} \int dp\ket{p}_{p_1}\ket{\eta\lambda_i^2 p}_{q_2}.\]
    This step encodes singular values of $A^TA$ into the entangled two qumodes.
    \item \emph{Regularization}. For a preset regularization parameter $\chi$ which is a hyperparameter, perform $e^{i\eta \chi \hat{p_1}\hat{p_2}}$ that shifts $q_2$ mode to $\ket{\eta(\lambda_i^2+\chi)p}_{q_2}$. Different values of $\chi$ can be tried that a specified $\chi$ can be chosen to get good enough  performance of prediction. 
    \item \emph{Singular-value transformation}. After a homodyne detection on qumode $q_2$ with result $q_2=0$, the state collapses to $\ket{\psi_{A^+}}$, where qumodes are discarded since they are disentangled from qubits. This step transforms quantum amplitudes from $\{\lambda_i\}$ to $\{\frac{\lambda_i}{\lambda_i^2+\chi}\}$.
    \item \emph{Prediction}. For new data $\tilde{\mathbf{a}}$, prepare the reference state $\ket{\Psi_R}=\ket{\mathbf{y}}\ket{\psi_{\tilde{\mathbf{a}}}}$. The prediction $\tilde{y}$ is proportional to the inner product between the target state $\ket{\psi_{\vec{A^+}}}$ and
    $\ket{\Psi_R}$, denoted as $\tilde{y}'=\braket{\psi_{\vec{A^+}}}{\Psi_R}$, which can be implemented using the method introduced in Refs.~\cite{rebentrost_14,cai_15}. The key point is to introduce an ancillary qubit to firstly construct an entangled state
    $\ket{\Psi}=\frac{1}{\sqrt{2}}(\ket{0}\ket{\Psi_R}+\ket{1}\ket{\psi_{A}})$. This can be achieved by setting the ancillary qubit in $(\ket{0}+\ket{1})/\sqrt{2}$ as a control state, and when the ancillary qubit is in $\ket{0}$ the reference state $\ket{\Psi_R}$ is prepared, and when in $\ket{1}$ state $\ket{\psi_{A}}$ is prepared. Then, conditioned on $\ket{1}$ state, procedures of $1-4$ for the algorithm are performed to generate $\ket{\psi_{A^+}}$, e.g., by attaching a control to all operations in procedures of $1-4$, which lead to a state $\ket{\Psi'}=\frac{1}{\sqrt{2}}(\ket{0}\ket{\Psi_R}+\ket{1}\ket{\psi_{A^+}})$. Then, a $\sigma_x$ measurement on the ancillary qubit of quantum state $\ket{\Psi'}$ gives the output $+1$ with probability $p=\frac{1}{2}(1+c\tilde{y}')$. Thus  $\tilde{y}'=\frac{2p-1}{c}$.
\end{enumerate}

The above-presumed infinite squeezing factor is impractical in
implementation, and success rate for post-selecting $q_s=0$ is
vanishing.

We now take finite squeezing into account, which leads to a modification of the algorithm in three places. Firstly, in the state preparation two qumodes are prepared as \[\ket{G_{12}}=s^{-1}\pi^{-\frac{1}{2}}\int dp_1dp_2 e^{-(p_1^2+p_2^2)/2s^2}\ket{p_1}_{p_1}\ket{p_2}_{p_2}\] with a squeezing factor $s$. Since $\ket{q}_q=\frac{1}{\sqrt{2\pi}}\int dpe^{-iqp}\ket{p}_p$, $\ket{G_{12}}=\ket{0}_{q_1}\ket{0}_{q_2}$ at $s\rightarrow\infty$ limit. Secondly, in the singular-value transformation, to fully disentangle qubits and qumodes, two qumodes are post-selected as $q_1=Q_1$ and $q_2=Q_2$ using homodyne detection. To get a nonzero success rate, we can let the point $(Q_1,Q_2)$ locates within a small area proportional to $\epsilon_q$ centered at $(0,0)$ (see Appendix~\ref{appendix:finite_squeezing}). Now, the unnormalized state can be written as
\begin{equation}
\ket{\psi_{A'}}=\sum_i  f_i(Q_1,Q_2)\frac{\lambda_i}{\lambda_i^2+\chi}\ket{\psi_{\mathbf{u}_i}}\ket{\psi_{\mathbf{v}_i}}
\end{equation}
where $f_i(Q_1,Q_2)\sim e^{-(Q_1^2+Q_2^2)/{2\alpha_i^2s^2}}$
at the limit $\alpha_i^2s^4\sim \varepsilon_q^{-1}$. Here $\alpha_i=\eta(\lambda_i^2+\chi)$. Compared with
$\ket{\psi_{A^+}}$, each coefficient now is multiplied by $f_i(Q_1,Q_2)$ correspondingly. When $\epsilon_q \rightarrow 0$, the state $\ket{\psi_{A'}}$ reduces to $\ket{\psi_{A^+}}$. In this sense $\epsilon_q$ characterizes an error. We may take $f_i(Q_1,Q_2)$ as an extra regularization due to finite squeezing, which will be discussed in Sec.~\ref{Sec: Discussion}. Thirdly, we use $\ket{\psi_{A'}}$ in the prediction. Note $\tilde{y}'$ is up to a factor to the required prediction $\tilde{y}$ as $\tilde{y}=c'\tilde{y}'$.  Such a factor can be obtained as the ratio between $y^{(m)}$ and $\tilde{y}^{(m)}$. Here $\tilde{y}^{(m)}$ is the predicted value of training data $\vec{x}^{(m)}$. The factor $c'$ can be more accurate by averaging on predicted values of a number of training data.

At this stage, we proceed to give some  implementation details.
While involving only with qumodes, standard techniques of Gaussian
quantum information processing~\cite{weedbrook_12} are sufficient
for the algorithm, which includes the following procedures. In the
state preparation, the squeezing state $\ket{G_{12}}$ can be
obtained using displacement operator and squeezing operator. In
the regularization step the operator is just a conditioned-phase
gate. Finally, in the singular value transformation, homodyne
detection of $q$ modes is required.

The essential part of the implementation  of the algorithm is to
construct the operator $U=e^{i\eta A^TA\hat{p_1}\hat{p_2}}$ for
the quantum phase estimation. As $U$ extracts singular values
implied in states of qubits and registers them on entangled
qumodes, it is essentially a hybrid  quantum operator. Effectively
constructing of $U$ involves a density matrix exponentization of
$A^TA$ using the method in Ref.~\cite{lloyd_14,kimmel_17}.
Further, the required exponential swap gate should be modified
\cite{lau_17} to couple the density matrix $A^TA$ to
$\hat{p_1}\hat{p_2}$. The density matrix exponentization $e^{i\rho
t}$ of density $\rho=A^TA$ can be obtained by repeatedly applies
the exponential swap operation and tracking out $\rho$,
\begin{eqnarray}\label{eq:exp_swap}
\text{Tr}_{\rho}(e^{i\delta t S_m}\rho\otimes\rho' e^{-i\delta t S_m})
= e^{i\delta t \rho}\rho' e^{-i\delta t \rho} + O(\delta t^2).
\end{eqnarray}
Here $S_m$ is the swap operator.
\begin{figure}
    \includegraphics[width=0.8\columnwidth]{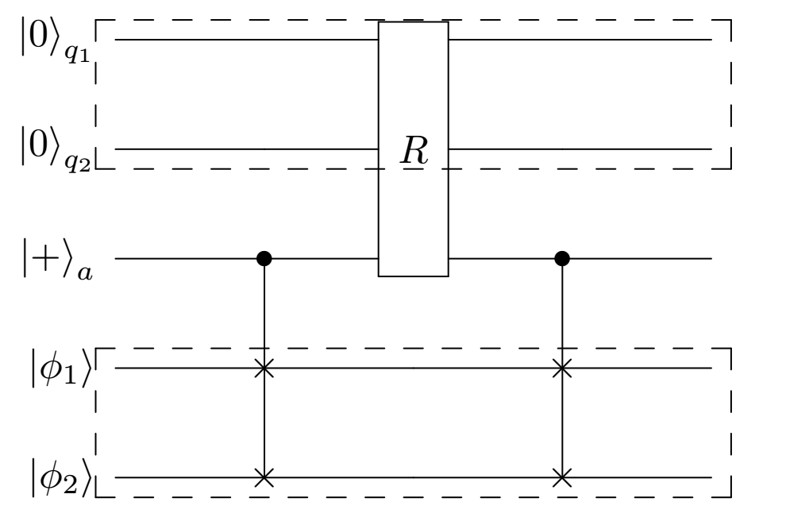}
    \caption{Quantum circuit for implementing modified exponential swap gate.
     An ancillary qubit is prepared at $\ket{+}$. Qumodes and qubits are not directly coupled.}
\end{figure}
To construct $U$, the exponential swap operator should be modified as
\begin{equation}\label{hybrid-swap}
e^{i\delta t \eta\hat{p_1}\hat{p_2} S_m}= C_SR^x_a
(\delta t \eta \hat{p_1}\hat{p_2})C_S.
\end{equation}
Here $R^x_a(\delta t \eta\hat{p_1}\hat{p_2})=e^{i\delta t \eta
\hat{p_1}\hat{p_2}\sigma_x}$, and $C_S= C_S^{bb'}C_S^{dd'}...$ is
a control swap operator on multi-modes, where
$C_S^{\alpha\alpha'}$ swap between $\alpha$ and $\alpha'$,
conditioned on the ancillary qubit $a$. Note that
Eq.~(\ref{hybrid-swap}) is a hybrid approach version of the one in
an optical system \cite{lau_17}. The conditional swap operator is
constructed as $C_S^{bb'}=C_{bb'}T_{ab'b}C_{bb'}$, where $C_{bb'}$
is a Control-NOT gate that takes $b$($b'$) as the control(target),
and $T_{ab'b}$ is the Toffoli gate that conditioned on both $b'$
and the ancillary qubit $a$. Initially the ancillary qubit $a$ is
set in state $\ket{+}_a=\frac{1}{\sqrt{2}}(\ket{0}_a+\ket{1}_a)$.
We now have
\begin{eqnarray*}
    &&\text{Tr}_{\rho}(e^{i\delta t \eta\hat{p_1}\hat{p_2} S_m}\rho\otimes\rho' e^{-i\delta t \eta\hat{p_1}\hat{p_2} S_m})\nonumber \\
    &&=
    1+i[\rho\hat{p_1}\hat{p_2},\rho']\eta\delta t +O(\delta t^2)\nonumber \\
    &&= e^{i\delta t\eta \rho\hat{p_1}\hat{p_2}}\rho' e^{-i\delta t\eta \rho\hat{p_1}\hat{p_2}} + O(\delta t^2).
\end{eqnarray*}
Set $\rho = A^TA$ we obtain $e^{i\delta t\eta A^TA\hat{p_1}\hat{p_2}}$.  To construct $U$ with desired precision $\varepsilon$ it requires $O(\varepsilon^{-1})$ copies of $\rho$\cite{lau_16}.

A key component of our hybrid approach  is to construct the hybrid operator
$R_a^x(\delta t \eta\hat{p_1}\hat{p_2})=e^{i \delta t \sigma_x\hat{p}_1\hat{p}_2}$ which involves a coupling of one qubit to two qumodes. Such a term would not appear naturally in a physical system. Nevertheless, it can be obtained by using basic hybrid quantum operators \cite{lloyd00}, with the following intertwined quantum evolution $e^{iH_2\delta t}e^{iH_1\delta t}e^{-iH_2\delta t}e^{-iH_1\delta t} =e^{-[H_1,H_2]\delta t^2}+O(\delta t^3)$.
Let $H_1=g\hat{p}_1\sigma_y$ and $H_2=g\hat{p}_2\sigma_z$ here and the evolution repeat ~$O(\frac{1}{g^2\delta t})$ times, we thus have $R_a(\delta t \eta\hat{p_1}\hat{p_2})$.

We briefly discuss how to implement the algorithm physically in
trapped ions, with some necessary modification. A more detailed demonstration can be found in Appendix.~\ref{appendix:trapped_ion}. It should be emphasized that the hybrid approach is more feasible for a physical implementation, comparing to the all-qubit
approach. For trapped ions,  the internal states and motional
states of trapped ions can respectively serve as discrete and
continuous variables for encoding information, and notably both
states are well-controllable in the trapped-ion systems
~\cite{cirac_95,lau_12,shen_14,ortiz-gutiérrez17,lamata_07,gerritsma_10}.
The setup contains ions in a Pauli trap lined in the $z$
direction. A specified ion, called a-ion, provides its motion
modes in $x$ and $y$ directions as two qumodes, and one qubit as
the ancillary qubit. We initialize two qumodes in squeezing states
of momentum, and the ancillary qubit in $\ket{0}_a$. Also, $n$
copies of $\ket{\psi_A}$ and the reference state $\ket{\psi_R}$ are
prepared using internal states of ions. In the phase estimation,
the conditional exponential swap operator can also be written as $e^{i\delta
t\eta\hat{q}_x\hat{q}_yS_m}=C_SH_aR^z_a(\delta
t\eta\hat{q}_x\hat{q}_y)H_aC_S$ for convenience in the present system of trapped ions, where $H_a$ is the Hadamard gate
on the ancillary qubit. The conditional swap operator $C_S$ has
been realized experimentally in trapped ions~\cite{linke_17}.
Setting $H_1 (\hat{q}_x)=g\hat{q}_x\sigma_x,H_2 (\hat{q}_y)=g\hat{q}_y\sigma_y$, which can
be realized using lasers for simulating Dirac
equations~\cite{lamata_07,gerritsma_10}, the hybrid gate
$R^z_a(\delta t\eta\hat{q}_x\hat{q}_y)$ is realizable, using the
intertwined quantum evolution of $H_1$ and $H_2$. The
regularization demands for a controlled-phase gate
$C_z=e^{i\chi\eta\hat{q}_x\hat{q}_y}$. Note that
$e^{ik_0t(\hat{q}_x+\hat{q}_y)^2}e^{-ik_0t\hat{q}_x^2}e^{-ik_0t\hat{q}_y^2}=e^{i2k_0t\hat{q}_x\hat{q}_y}$.
Then by relaxing the trap along $x,y$ directions and strengthening
it along $x=-y$ direction for a-ion, we can realize $C_z$
approximately by setting $2k_0t=\chi\eta$. The singular value
transformation can be implemented with homodyne detection of
approximate zero momentum, using the motion tomography method in
Ref.~\cite{poyatos_96}. In the prediction, a swap
test~\cite{buhrman_01} can access $|\tilde{y}|=\sqrt{2p-1}$, where
$p=\frac{1}{2}(1+|\braket{\psi_{A^+}}{\psi_R}|^2)$ is the success
rate of projecting the ancillary qubit into $\ket{0}_a$ state. The
sign of $\tilde{y}$ should be determined by other
means~\cite{schuld_16}. 

\section{Discussions and conclusions}~\label{Sec: Discussion}
The hybrid approach for quantum linear regression is similar to that of the all-qubit approach. Both use auxiliary modes to encode information of singular values; and finally project out the auxiliary modes which at the same time gives the singular value transformation. However, the hybrid approach exploits the infinite dimension of qumodes and is essentially different in several aspects. Firstly, it uses entangled qumodes to achieve the singular value transformation. This is unique for continuous variable quantum states (see Appendix~\ref{appendix: qumode} for a more general investigation). Secondly, the finite squeezing factor $s$ plays a basic role which is absent for the all-qubit approach. Those lead to some distinct properties for the algorithm, as would be revealed below.

We first analyze the runtime of the hybrid approach using two qumodes. It mainly arises from the construction of operator $U=e^{i\eta A^TA\hat{p_1}\hat{p_2}}$ for the quantum phase estimation, and the success rate of homodyne detection at the step of singular value transformation. To construct $U$,  it  requires $O(\varepsilon^{-1})$ copies of density matrix $A^TA$ with desired precision $\varepsilon$. Meanwhile, each copy of density matrix $A^TA$ can be obtained from $\ket{\psi_A}$ that takes runtime $O(\log{(MN)})$ using qRAM.   Thus the runtime in the phase estimation is $O(\varepsilon^{-1}\log{(MN)})$. The runtime for singular value transformation turns to be $O(\varepsilon_q^{-\frac{3}{2}})$ (see Appendix~\ref{appendix:finite_squeezing}). In total, the runtime scales as $O(\varepsilon_q^{-\frac{3}{2}}\epsilon^{-1}\log{(MN)})$.

We investigate the role of the regularization parameter $\chi$ and the squeezing factor $s$ for quantum speed-up and accuracy. For low-rank $A^TA$, we expect $\lambda_{\text{min}}\sim O(1)$ and $\lambda_{\text{min}}\sim O(N^{-{\frac{1}{2}}})$, which corresponds to the case of bad conditional number.
Without regularization, the exponential speed-up loses in the all-qubit approach~\cite{rebentrost_14}. In our hybrid approach, however, exponential speed-up always holds under a constant error $\epsilon_q$ once we set $s\sim O(N^{{\frac{1}{2}}})$ (recalling that $(\lambda_i^2+\chi)^2s^4\sim\epsilon_q^{-1}$). Moreover, with regularization, the required squeezing factor becomes $s\sim O(\sqrt{\log(N)})$ if we set $\chi \sim O(\log N)$, which greatly reduces the requiring resource of squeezing. For a preset constant $\chi$, the requiring squeezing factor turns to be $s\propto \frac{1}{\sqrt{\chi}}$. 

For the aspect of experimental reachable squeezing factor, the squeezing of motional modes of trapped ions can reach about $12.6$ dB~\cite{kienzler_15} (a squeezing factor about $s\sim 18$), corresponding to four qubits for encoding a continuous number. The precision can be enough for our demonstration proposal of quantum linear regression with trapped ions where only two singular values need to be told apart (See Appendix~\ref{appendix:trapped_ion} where we have used a data set with four samples and two features). The squeezing factor can be even higher, which may be raised, e.g., with a longer parametric drive duration time when keeping coherence~\cite{burd_18}. We may see that raising the squeezing factor requires an improvement of operations on a single qumode against the noisy environment~\cite{kienzler_15,burd_18}. Such a challenge is different from that of increasing the number of qubits, which depend on the scalability of physical setups. 


Due to the
regularization and finite squeezing, the obtained state $\ket{\psi'}$ is different from one with infinite squeezing and without regularization. To characterize this difference, we calculate the fidelity between $\ket{\psi'}$ and $\ket{\psi_{A}^+}$ and study its relation to regularization factor $\chi$ and squeezing factor $s$. More details can be found in Appendix.~\ref{appendix:error}. As expected,
the fidelity increases when $s$ increases
as one should expect, and decreases when $\chi$ increases. Moreover,
under larger regularization the fidelity can decrease significantly slower with reducing squeezing factor $s$.

We further mention the shrinkage effects~\cite{Hansen1987} due to the
regularization and finite squeezing, which may benefit machine learning. Firstly, regularization
introduces a shrinkage that the coefficients  in
$\ket{\psi_{A^+}}$ are $\frac{\lambda_i}{\lambda_i^2+\chi}$,
instead of $\frac{1}{\lambda}$ for the case without
regularization. The extent of shrinkage is higher for small
singular values (low variance components) and lower for large
singular values (high variance components). Further, in the case
of finite squeezing, the final state we obtain is
$\ket{\psi_{A'}}$ instead of $\ket{\psi_{A^+}}$, that is, each
coefficient is rescaled by $f_i(Q_1,Q_2)$, correspondingly. This
further reduces the weighting of low variance components. Rather
than taking $f_i(Q_1,Q_2)$ as an imperfection of the algorithm, we
can consider it as an extra regularization due to finite
squeezing. Whether this regularization will improve the
performance of prediction is an interesting question and we would
leave it for further investigation.

It is  meaningful to highlight the role of squeezing factor $s$ in
the algorithm. On one hand, it is related to the success rate of
post-selection. The smaller $s$, the bigger $\epsilon_q$, and thus
the higher the success rate. On the other hand, smaller $s$ leads
to a bigger shrinkage. Thus, a proper squeezing factor $s$ is not
only necessary for efficiently implementing the quantum algorithm,
but also may be beneficial for machine learning.

In summary, we have adopted a hybrid approach for quantum linear
regression, exploiting the advantage  of qubits for encoding data
and the efficiency of auxiliary qumodes for implementing phase
estimations and singular value transformation. The regularization
can be incorporated directly with a controlled phase gate on two
qumodes. This regularization can remedy the issue of bad condition
number by reducing the requirement of squeezing resource, and thus
is important for running the quantum algorithm efficiently. Our
hybrid approach has the same order of runtime as the all-qubit
regarding to the dimension of data $N$, the number of training
data (samples) $M$, and the precision $\varepsilon$, but can save
the using of ancillary qubits at the price of requiring finite
squeezing states with qumodes. We have also demonstrated the
important role of finite squeezing for efficiently running the
algorithm. Moreover, we have shown that the shrinkage effect due
to finite squeezing may provide a new type of regularization for
linear regression. We wish that the hybrid approach may allow us
to design quantum algorithms with more flexibility and efficiency,
and may even provide new insights for quantum machine learning, by
exploiting the infinite dimensionality as well as the finite
squeezing nature of qumodes.


\acknowledgments{We thank Mile Gu for helpful discussions. This
	work was supported by the NKRDP of China (Grant
	No. 2016YFA0301800) and the NSFC (Grants No. 91636218
	and No. 11474153) as well as the KPST of Guangzhou (Grant
	No. 201804020055)}

\appendix{
	\section{Qumodes for quantum computing} \label{appendix: qumode}
	
	\emph{Quantum phase estimation}. At the heart of QPE is quantum Fourier transformation (QFT) which gives
	exponential speed-up, as performing QFT on $N$ qubits takes the
	order of $N^2$ quantum operations, while fast Fourier
	transformation takes $O(N2^N)$. It is noted that Fourier
	transformation for quadrature field operators $\hat{p}$ and
	$\hat{q}$ is inherent, as revealed by the relation of their
	eigenstates $\ket{q}$ and $\ket{p}$, which takes the form
	$\ket{q}=\frac{1}{\sqrt{2\pi}}\int dpe^{-iqp}\ket{p}$. Based on
	the inherent QFT in qumodes, quantum phase
	estimation~\cite{lloyd00} routine can be much simplified. Note
	$H\ket{e_i}=E_i\ket{e_i}$. To write eigenvalues into the
	registering qumode, one just needs to performs quantum gate
	$U=e^{iH\hat{p}}$ on $\ket{\psi}\ket{0}_q$, which results in
	$\sum_i \psi_i\ket{e_i}\ket{E_i}_q$.  The registering of
	eigenvalues in qumodes is just a shift of position (momentum),
	with the quantity determined by the eigenvalues obtained as $H$
	works on its eigenstate. Remarkably, only a single qumode with
	squeezing factor~\cite{liu_16} $\varepsilon^{-1}$  is required for
	desired precision $\varepsilon$.  In contrast, for all-qubit
	system, registering $E_i$ with same precision requires
	$O(\log{\varepsilon^{-1}})$ qubits.
	
	\emph{Quantum addition}.  We consider the addition $a+b$. First encode $a$ as $\ket{a}_q$, then the addition can be realized by performing the shift operator $e^{ib\hat{p}}$ on $\ket{a}_q$, and it is easy to see that the output is $\ket{a+b}_q$. In general, floating computing is more conveniently implemented in the continuous-variable quantum computing setup~\cite{lloyd_99}.
	
	\emph{Transformation of quantum amplitudes using entangled qumodes}.
	The task is to transform the initial state $\sum_i\ket{\psi_i}$ to $\sum_ig(\lambda_i)\ket{\psi_i}$ (unnormalized), where $\lambda_i$ relates to $\ket{\psi_i}$, such as a pair of eigenvalue and eigenstate. Here quantum amplitudes are mapped as $a_i\rightarrow a_ig(\lambda_i)$. We show how this can be achieved using entangled qumodes and projection. This is unique for entangled states of continuous-variable. For illustration we give the derivation only for infinite squeezing.  We assume that the following quantum state can be prepared,
	\begin{equation}
		\ket{\psi}=\sum_i \ket{\psi_i}\int dp\ket{p}_{p_1}\ket{\lambda_i p}_{q_2}.
	\end{equation}
	Take a homodyne detection on qumode  $q_2$ and if the result is $\lambda_i p= q_s$,  then the state collapses to
	\begin{equation}
		\sum_i \frac{1}{\lambda_i} \ket{\psi_i}\ket{\frac{q_s}{\lambda_i}}_{p_1},
	\end{equation}
	where $q_2$ mode has been discarded. Further, post-selecting of $q_s=0$, then this can realize $g(\lambda)=\frac{1}{\lambda}$.  General $g(\lambda)$ in principle can be implemented as follows. Firstly, quantum float computation~\cite{lloyd_99} allows a mapping from $\ket{\lambda_i p}_{q_2}$ to $\ket{1/g(\lambda_i p)}_{q_2}$. Then, a homodyne measurement on the qumode $q_2$ leads to the state
	\begin{equation}
		\sum_i g(\lambda_i) \ket{\psi_i}\ket{\frac{g^{-1}(q_s^{-1})}{\lambda_i}}_{p_1},
	\end{equation}
	which multiples the coefficient by $g(\lambda_i)$, respectively. It is required that $g^{-1}(q_s^{-1})=0$ for some given $q_s$ to disentangle the $p_1$ qumode. For instance, when $g(x)=x^{-n}$ ($n>0$), we can chose $q_s=0$.
	
	~~~~~~~
	\section{Finite squeezing analysis}
	\label{appendix:finite_squeezing}
	Let us consider squeezed states with squeezing factor $s$, following the method in Ref.~\cite{lau_17}. Then, the unitary operator $U=e^{i\eta A^TA\hat{p_1}\hat{p_2}}$ performs on the prepared state
	\begin{equation}
		\ket{\Psi} \propto \ket{\psi_A}\int dp_1dp_2 s^{-1}e^{-(p_1^2+p_2^2)/2s^2}\ket{p_1}_{p_1}\ket{p_2}_{p_2}.
	\end{equation}
	We emphasize that the factor $s^{-1}$ should not be dropped when analyzing the runtime behavior. It follows by the operator $e^{i\eta \chi \hat{p_1}\hat{p_2}}$ realizing a regularization.
	To unentangle qumodes from qubits, homodyne detections are conducted on both qumodes with results $q_1=Q_1$ and $q_2=Q_2$. Then the quantum state turns to be
	\begin{equation}
		\sum_i \lambda_i B_i(Q_1,Q_2)\ket{\psi_{\mathbf{u}_i}}\ket{\psi_{\mathbf{v}_i}}\ket{Q_1}_{q_1}\ket{Q_2}_{q_2},
	\end{equation}
	where
	\begin{align}
		B_i(Q_1,Q_2)= \int dp_1dp_2 e^{-(p_1^2+p_2^2)/{2s^2}}e^{i(\alpha_ip_1p_2-p_1Q_1-p_2Q_2)} \nonumber\\
		\propto \frac{\exp(-\left[s^2(Q_1^2+Q_2^2)+2is^4\alpha_iQ_1Q_2\right]/{2(1+s^4\alpha_i^2)})}{s\alpha_i\sqrt{1+1/{s^4\alpha_i^2}}}.
	\end{align}
	For brevity we have introduced $\alpha_i=\eta(\lambda_i^2+\chi)$.
	Set $\alpha_i^2s^4\sim \varepsilon_q^{-1}$, then
	\begin{equation}
		B_i(Q_1,Q_2) \sim \frac{e^{-(Q_1^2+Q_2^2)/{2\alpha_i^2s^2}}}{s\alpha_i}.
	\end{equation}
	The width of distribution of both $Q_1$ and $Q_2$ is $\alpha_is \sim \frac{1}{s\sqrt{\varepsilon_q}}$. For our task  it is natural to let the precision $\varepsilon_q$ specified to $\alpha_i$. We let $\frac{Q_1^2+Q_2^2}{\alpha_i} \lesssim \varepsilon_q$. Note the probability density is $|\lambda_iB_i(Q_1,Q_2)|^2$. Then,
	the success rate of homodyne detection of two qumodes around the center with area $\varepsilon_q$ can be approximately calculated through
	\begin{align}
		& \sum_i\int_{Q_1^2+Q_2^2\leq\alpha_i\varepsilon_q} dQ_1dQ_2|\lambda_iB_i(Q_1,Q_2)|^2 \nonumber\\
		&=\sum_i\lambda_i^2\int_{0}^{\sqrt{\varepsilon_q\alpha_i}}\frac{e^{-r^2/{\alpha_i^2s^2 }}}{s^2\alpha_i^2}rdr \propto \varepsilon_q^{\frac{3}{2}}
	\end{align}
	
	The runtime is $O(\varepsilon_q^{-\frac{3}{2}})$, and further if amplitude amplification~\cite{brassard_00} can be applied to continuous variables then the runtime would reduce to $O(\varepsilon_q^{-\frac{3}{4}})$.

	\section{Physical implementation: a demonstration}\label{appendix:trapped_ion}
	For the purpose of demonstration, we consider a dataset that contains four training samples, each having two features. This corresponds to $M=4$ and $N=2$. Features for each sample is represented by $\vec{a}^{(m)}=(a_0^{(m)},a_1^{(m)})$ ($m=0,1,2,3$), and the corresponding target value is $y^{(m)}$.  The regression task is to predict the target value for new data $\tilde{\vec{a}}=(\tilde{a}_0,\tilde{a}_1)$.  
	
	We choose trapped ions~\cite{cirac_95,lau_12,shen_14,ortiz-gutiérrez17,lamata_07,gerritsma_10} for a physical implementation of the algorithm. Our setup contains ions in a Pauli trap lined in the $z$ direction. Each ion can provide a qubit and two qumodes (motion modes in $x$ and $y$ directions). 
	According to their roles in the algorithm, we can group them as following:
	\begin{enumerate}
		\item d-ions contains three ions for loading the data information into a quantum state $\ket{\psi_A}=\sum_{m=0}^{3}\sum_{n=0}^{1}a_n^{(m)}\ket{m}\ket{n}$, which would be transformed into the target state $\ket{\psi_{A^+}}$.
		\item t-ions contains three ions. They are initialized to $\ket{\psi_{A}}$ and help to construct $U$ for the phase estimation. Many copies of t-ions may be required depending on the desired precision. Here we simply choose two copies for the purpose of demonstration.	 
		\item a-ion provides an ancillary qubit and two qumodes. 
		\item r-ions contains three ions for encoding the reference state $\ket{\Psi_R}$.  	
	\end{enumerate} 
	
	In total thirteen ions are required.
	We briefly discuss how those ions play their role in the procedures of the algorithm, as illustrated in Fig.~{\ref{active_ions}}. Firstly, a-ion, d-ions, t-ions and r-ions are initialized to $\ket{0}_a\ket{G_{12}}$, $\ket{\psi_A}$,$\ket{\psi_A}$, $\ket{\psi_R}$, respectively. In the training step, which transfers $\ket{\psi_A}$ to $\ket{\psi_{A^+}}$, t-ions, d-ions, and a-ion are involved to perform the quantum phase estimation. For either t-ions or d-ions, only the qubit recording features is involved, as marked red in Fig.~{\ref{active_ions}}. Then, a controlled-phase gate and homodyne measurements performs on two qumodes of the a-ion, realizing regularization and singular value transformation respectively. In the prediction, a swap test involves r-ions (encoding $\ket{\Psi_R}$), d-ions (encoding $\ket{\Psi_A^+}$), and a-ion (providing an ancillary qubit). Note that two motion modes of a-ion are not required in this stage anymore. 
	
	\begin{figure}
		\includegraphics[width=0.8\columnwidth]{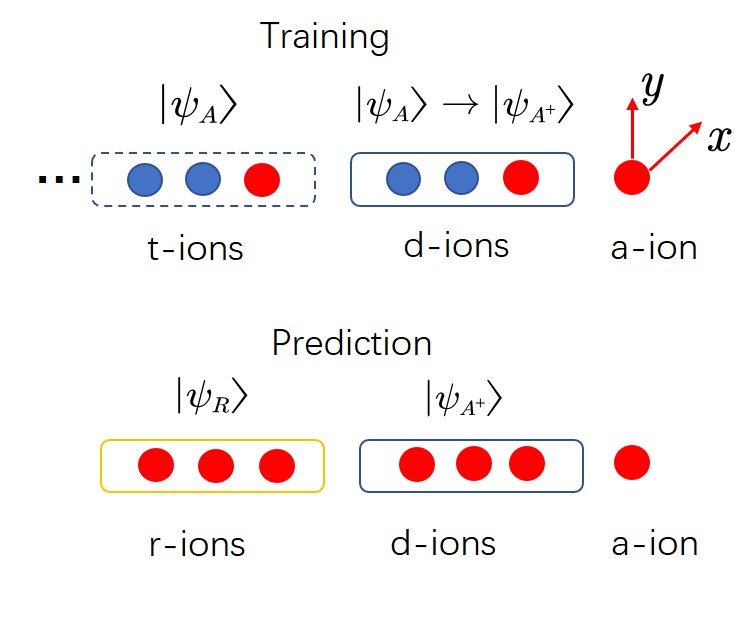}
		\caption{Ions (in red color) involved at the training and prediction steps.}\label{active_ions}
	\end{figure}
	We work on squeezing states of momentum, as a regularization operator $e^{i\eta\chi\hat{x}\hat{y}}$ is more realizable in a single ion. 
	The implementation details with the five procedures are given as follows: 
	\begin{enumerate}
		\item \emph{State preparation}. d-ions are prepared in the state by $D_A\ket{0}_{d_1}\ket{0}_{d_2}\ket{0}_{d_3}=\ket{\psi_{A}}$. Explicit quantum circuit of $D_A$ can be seen in Fig.~\ref{state_paration}. t-ions are also prepared in state $\ket{\psi_{A}}$. For the a-ion, the qubit state is initialed as $\ket{0}_a$, and the motion state is initialized as $\ket{G_{12}}=\ket{0,s}_{p_x}\ket{0,s}_{p_y}$, where squeezing state $\ket{0,s}_{p_\alpha}$ is obtained by performing a squeezing operator $S_{\alpha}(1/s)=e^{\frac{i}{2}\ln{s}(\hat{q}_\alpha\hat{p}_\alpha+\hat{p}_\alpha\hat{q}_\alpha)}$ on the vacuum state $\ket{0}_{c_\alpha}$. r-ions are initialized in the reference state $\ket{\psi_R}=D_{\vec{y}}\otimes D_{\tilde{\vec{a}}} \ket{0_10_2}_{r}\ket{0}_{r_3}=\ket{\psi_{\mathbf{y}}}\ket{\psi_{\tilde{\mathbf{a}}}}$. Here $\ket{0_10_2}_{r}$ short-notes $\ket{0}_{r_1}\ket{0}_{r_2}$. $D_{\vec{y}}$ is a two-qubit operator that prepares $\ket{\psi_{\mathbf{y}}}=y^{(0)}\ket{0_10_2}_{r}+y^{(1)}\ket{1_10_2}_{r}+y^{(2)}\ket{0_11_2}_{r}+y^{(3)}\ket{1_11_2}_{r}$, and $D_{\tilde{\vec{a}}}\ket{0}_{r_3}=\tilde{a}_0\ket{0}_{r_3}+\tilde{a}_1\ket{1}_{r_3}$.
		\begin{figure}
			\includegraphics[width=0.8\columnwidth]{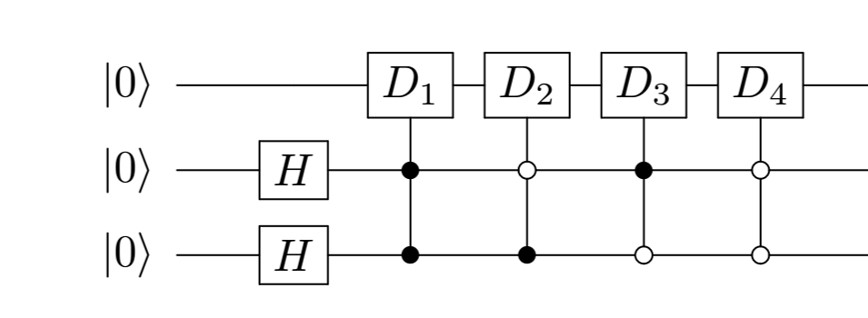}
			\caption{Quantum circuit for preparing $\ket{\psi_A}$. Here, $D_m=e^{i\theta_m\sigma_y}$ ($m=0,1,2,3$) with $\theta=\arctan(a^{(m)}_1/a^{(m)}_0)$ are single-qubit gates, which realize $D_m\ket{0}=a^{(m)}_0\ket{0}+a^{(m)}_1\ket{1}$. $H$ is the Hadamard gate. Two controls are added to $D_m$, such that they generate $\ket{m}(a^{(m)}_1\ket{0}+a^{(m)}\ket{1})$, respectively. Here we have defined $\ket{0}\equiv \ket{00}$, $\ket{1}\equiv \ket{10}$, $\ket{2}\equiv \ket{01}$, $\ket{3}\equiv \ket{11}$. The quantum circuit outputs the required state $\ket{\psi_A}=\sum_{m=0}^{3}\sum_{n=0}^{1}a_n^{(m)}\ket{m}\ket{n}$.}\label{state_preparation}
		\end{figure}
		
		\item \emph{Quantum phase estimation}. This step applies an unitary operation $U=e^{i\eta A^TA\hat{q_x}\hat{q_y}}$, using $\text{Tr}_{\rho}(e^{i\delta t \eta\hat{q}_x\hat{q}_y S_m}\rho\otimes\rho' e^{-i\delta t\eta\hat{q}_x\hat{q}_y S_m})$. Here $\rho\equiv A^TA=\text{Tr}_{t_1t_2}\ket{\psi_A}\bra{\psi_A}$ is a state on t-ions, and $\rho'\equiv\ket{\psi_A}\bra{\psi_A}$ is a state
		on d-ions.  The conditional exponential swap operator $e^{i\delta t\eta\hat{q}_x\hat{q}_yS}$ is $C_SH_aR^z_a(\delta t\eta\hat{q}_x\hat{q}_y)H_aC_S$. Here $C_S$ is a conditional swap operator performing on $t_3$ qubit and $d_3$ qubit, conditioned on the qubit  of a-ion. We emphasize that recently the conditional swap operator has been realized experimentally in trapped ions~\cite{linke_17}. $R^z_a(\delta t\eta\hat{q}_x\hat{q}_y)=e^{i\delta t\eta\sigma_z\hat{q}_x\hat{q}_y}$ performs on the a-ion. Since $H_1=g\hat{q}_x\sigma_x$ and $H_2=g\hat{q}_y\sigma_y$ can be realized using lasers when simulating Dirac equations~\cite{lamata_07,gerritsma_10}, then a quantum evolution $e^{iH_2\delta t}e^{iH_1\delta t}e^{-iH_2\delta t}e^{-iH_1\delta t}$ can realize $R^z_h(\delta t\eta\hat{q}_x\hat{q}_y)$ by repeating this evolution $\eta/(g^2\delta t)$ times.
		
		
		\item \emph{Regularization}. The regularization operator is a controlled-phase gate $C_z=e^{i\chi\eta\hat{q}_x\hat{q}_y}$. This may be obtained using the relation: $e^{ik_0t(\hat{q}_x+\hat{q}_y)^2}e^{-ik_0t\hat{q}_x^2}e^{-ik_0t\hat{q}_y^2}=e^{i2k_0t\hat{q}_x\hat{q}_y}$ by setting $2k_0t=\chi\eta$. Physically, it can be realized by disturbing the trap for the a-ion as follows. First relax the trap both along $x$ and $y$ direction by $k_0$ in a short time period $t$, then enhance the trap along $x=-y$ direction by $k_0$ in a same period $t$. If $t$ is small enough, this evolution can approximate $C_z$ once $k_0=\chi\eta/t$.  
		
		\item \emph{Singular-value transformation}. We need a measurement that projects the momentum of a-ion in an small area $\varepsilon_q$ around zero point.
		A modification of the motion tomography~\cite{poyatos_96} can be applied for this task. The projective measurement is  $\Pi_a(s)=S^+(1/s)\Pi_a(0)S(1/s)$, where $\Pi_a(0)$ project to the ion's motional vacuum state $\ket{0}_c$. To achieve this, we first apply $S(1/s)$ both in $x$ and $y$ directions, then use the quantum jump method to get the probability of successful projection $\Pi_a(s)$. A projection onto a small area around zero momentum may be approximated by choosing a slightly bigger squeezing factor, which raises the success rate.   	
		
		\item \emph{Prediction}. The swap test requires a conditional swap operation, which swaps states between d-ions and r-ions, conditioned on the qubit in a-ion.  The qubit in a-ion should be reseted in $\ket{+}_a$ state. After the swap operation, a Hadamard gate performs on the qubit of a-ion, and project it onto $\ket{0}_a$ state. The probability of successful projection is $p=\frac{1}{2}(1+|\braket{\psi_{A^+}}{\psi_R}|^2)$, and we have $|\tilde{y}|=\sqrt{2p-1}$. The sign of $\tilde{y}$ should be determined by other means. 
	\end{enumerate}
}

\section{Error due to finite squeezing and regularization}\label{appendix:error}
\begin{figure}\label{fidelity_to_s_chi}
	\includegraphics[width=0.8\columnwidth]{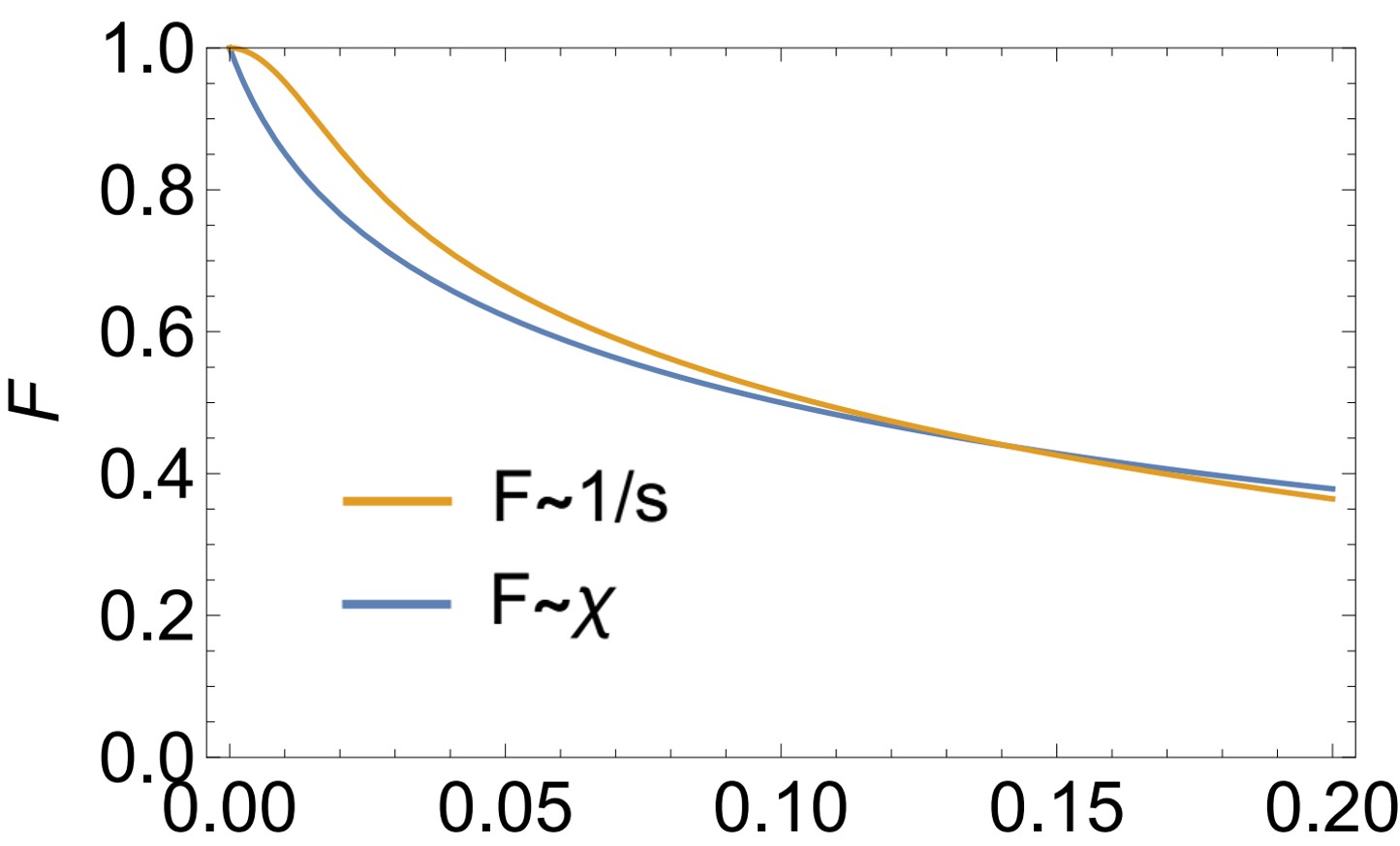}
	\caption{The relation of fidelity with squeezing $\frac{1}{s}$ (in brown) and with the regularization $\chi$ (in blue). }
\end{figure}
We analyze the effect of finite squeezing and regularization, by calculating the fidelity between the 
$\ket{\psi_{A_0^+}}=\sum_i\frac{1}{\lambda_i}\ket{\psi_{\mathbf{u}_i}}\ket{\psi_{\mathbf{v}_i}}$ and $\ket{\psi_{A'}}$. The fidelity turns to be $    F(s,\chi) \sim \sum_i f_i(Q_1,Q_2)\frac{1}{\alpha_i}$ (not normalized). Without loss of generality, we assuming $\lambda_i^2$ distribute uniformly in the zone $[\delta_0,1]$, where $\delta_0$ is a small positive number to avoid divergence brought by $1/\lambda$. Then the fidelity can be approximated as 
\begin{eqnarray}
	F(s,\chi) &\propto& \int_{\delta_0}^{1}d\lambda \exp(-\frac{\epsilon_q}{2\eta^2(\lambda+\chi)^2s^2})\frac{1}{\eta(\lambda+\chi)} \nonumber \\
	&=&E_i(-\frac{\epsilon_q}{2(\delta_0+\chi)^2s^2\eta^2})-E_i(-\frac{\epsilon_q}{2(1+\chi)^2s^2\eta^2})\nonumber \\
\end{eqnarray}
where $E_i(z)=-\int_{-z}^{\infty}dte^{-t^2}/t$.
The fidelity increases when $s$ increases
as one should expect, and decrease when $\chi$ increase, as seen in Fig.~3.

We also fixed the regularization, and study the behavior of fidelity with $\frac{1}{s}$ between $\ket{\psi_A^+}$ and $\ket{\psi_A'}$ under different $\chi$. As seen in Fig.~4, with larger regularization the fidelity can decrease slower when reducing the squeezing factor $s$.

\begin{figure}\label{fidelity_to_s}
	\includegraphics[width=0.8\columnwidth]{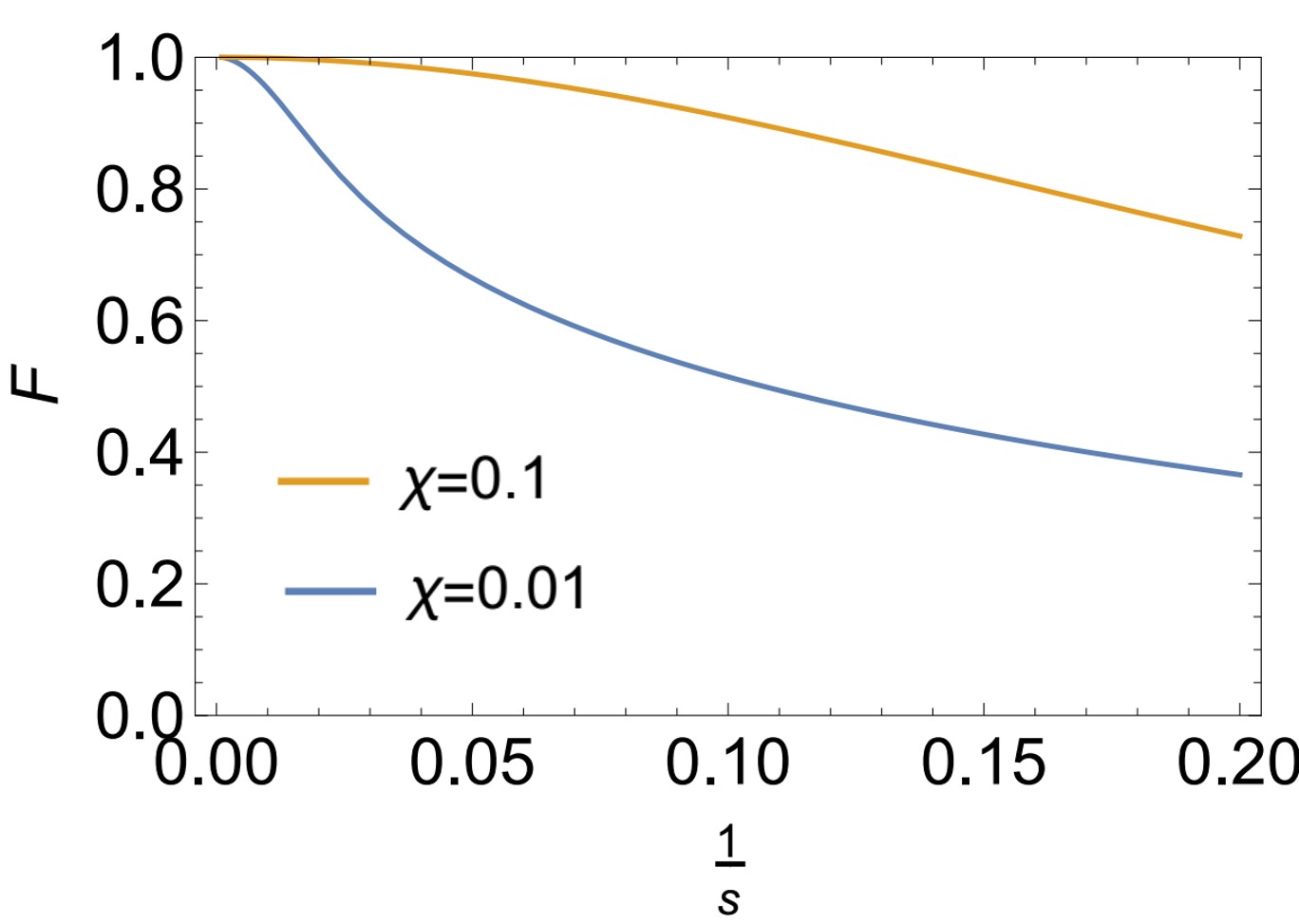}
	\caption{Fidelity with squeezing under different regularization $\chi=0.01,0.1$.}
\end{figure}

%

\end{document}